# Not even the air of empty spaces is coronavirus free
# (Two meters is not a safe distance)

Edilson Crema[1]

*Nuclear Physics Department, University of São Paulo, Brazil*

**Abstract**

At the moment, several countries are coming out of social confinement while others are still in the midst of the severe acute respiratory syndrome coronavirus 2 (SARS-CoV-2) outbreak. In both cases, a key safety measure encouraged by health authorities is the distance of one to two meters between people. This recommendation created in the population the false idea that, by staying two meters from each other, it is not necessary to use a mask or other protections. Even the highest leaders of the World Health Organization conduct daily interviews without a mask. This recommended two-meters distance is mainly based on short-distance contagion, when infected drops are exhaled during a speech, coughing, or sneezing and directly hit another person. The dangerous form of airway contamination caused by droplets that remain suspended in the air for several hours has been almost ignored. However, the theoretical calculations performed in this work, recent experiments, and the accumulated knowledge in this and other epidemics reveal that, because of the airborne transmission, there is no safe distance to the coronavirus, either indoors or in open places. Recent investigations have confirmed not only the presence of the coronavirus in droplets suspended in the air but that these viruses remain active for several hours. Furthermore, significant indirect evidence of this means of transmission is the great difference in contagion between Brazilian regions in the current outbreak. While the Amazonian states have a contamination rate greater than 20%, in the southern states of the country this rate is less than 1%, despite high temperatures in the Northern region. Notwithstanding the social and economic differences between these regions, it seems that the extremely high humidity of the forest air prolongs the survival of the viruses in the drops in the external environment. Our theoretical calculations explain empirical observations from recent epidemiological studies and strengthen the need to use, not only a mask but also protective glasses throughout the population in the same way that they are mandatory for health professionals. Besides, our calculations show how air conditioning and heating systems can increase contagion. Finally, we suggest measures that could reduce the spread of the pandemic.

**Introduction**

The ray of light that runs through a dark room reveals the existence of numerous small grains of dust that can float in the air for a long time. Since antiquity, this phenomenon was already known. A famous observation of this effect is documented in Lucretius's poem De Rerum Natura, written around 50 BC. In addition to the empirical description of the phenomenon, and following the tradition of Democritus and Epicurus,

---

[1] crema@if.usp.br

Lucretius also proposed an atomistic explanation for the support of particles in the air, according to which their weight would be compensated by the collisions of air atoms [1]. However, the behavior of tiny bodies immersed in fluids was only understood from the 19th century on owing to the works of Robert Brown [2], George Gabriel Stokes, and finally with Einstein's famous work of 1905, *On the movement of small particles in suspension within liquids at rest*. Currently, this phenomenon has gained tragic relevance due to the uncontrolled dispersion of the Covid-19 throughout the planet, since airborne transmission is one of the forms of viral contamination, as well as the direct reception of drops exhaled by a contaminated person and contact with infected surfaces. There is still no consensus among researchers as to which of these forms of contagion is the most important in the case of the coronavirus. Even so, at the end of March 2020, the World Health Organization released a bulletin stating that there was insufficient scientific evidence that SARS-COV-2 was significantly airborne transmitted. This interpretation, which minimized the importance of this form of contamination, has prevailed since the beginning of the pandemic, leading the World Health Organization and several governments to insist that the use of masks was unnecessary throughout the population. However, with the rapid spread of the virus in countries and in the world, the deadly reality has imposed itself and forced the planetary health authorities to reverse this directive, saving thousands of lives by requiring the use of masks in several countries. From a scientific point of view, this late change in positioning was the authorities' recognition of that air transmission of SARS-COV-2 is an unquestionable fact. Nevertheless, it remains to be understood how this process takes place. In this article we will try to help clarify the physical processes involved in this means of contamination and we will show that there are still some important recommendations that health authorities should indicate to reduce viral transmissibility.

Despite being the third outbreak of a coronavirus in less than two decades, existing research had not yet fully understood the transmission mechanisms of this virus. A similar situation occurred with the Influenza virus. While some important books and works drew attention to the relevance of its transmission by aerosols (droplets) [3-6], other authors argued that short-distance transmission by drops would be the main means of infection [7,8], and this latter position prevailed for a long time among health authorities [9,10] who practically ignored airborne transmission. Perhaps this is why, at the beginning of the current COVID-19 pandemic, the most important health authorities on the planet recommended that only hand washing and a distance of two meters would be safe protection procedures. Neither was the widespread use of masks recommended.

However, the transmission of the coronavirus proved to be much more complex and dangerous, and the airway may have had played a decisive role in this process. One of the hypotheses raised to explain its rapid spread is the fact that this virus strongly attacks the deep respiratory system and lungs causing the drops expelled by coughing and sneezing to carry higher viral doses. Important work was published recently, where a video unequivocally demonstrates that, even during a person's normal speech, hundreds of drops of saliva and/or secretions are produced and projected into the air, and large drops can remain in suspension for several minutes [11,12]. On the other hand, sneezing and coughing expel thousands of drops, which are of various sizes and are composed of saliva and/or mucous secretions from the respiratory system, constituting an important means of transporting viruses to the external environment [13]. Analysis of the most important focus of coronavirus transmission in the 2002-2003 epidemic in Hong Kong, the Amoy Gardens residential complex, demonstrated that a single infected person contaminated, through the ventilation system, more than 300 people living above him in the same building. In addition, there are also strong indications that, in the same Amoy Gardens, contaminant droplets could have been carried by the wind for several tens of meters and infected people in another building, far from patient zero [14]. Also, during the coronavirus (MERS-COV) epidemic in South Korea in 2015, research conducted in two hospitals that were sources of contamination revealed that transmission through ambient air may have been one of the main means of contagion [15].

Concerning the current outbreak, recent simulation demonstrated that the likely onset of the COVID-19 pandemic in Guangzhou, China, occurred by air in a restaurant, where the air conditioning system played a decisive role in the spread of viruses exuded by an infected man who had just arrived from Wuhan [16]. Another important experimental finding of the possible relevance of micro drops (aerosols) in aerial transmission has just been disclosed by Kyoto University, in which the movement of drops expelled during coughing or speech were also filmed with laser beams, revealing that while larger drops fall rapidly and settle on the ground and furniture, there are hundreds of micro drops that remain suspended in the air several hours after being expelled. And, most seriously, these small drops disperse rapidly and, a few minutes after their production, occupy the entire environment, covering distances greater than 8 meters [17]. These experimental results call into question one of the main recommendations of health authorities to contain the outbreak: the distance of 1m to 2m between people would be a safe method of prevention. This indication is based only on transmission by larger drops, dangerously ignoring the airway. With this, the false idea was created in the population that, by staying two meters away from other people, it is not necessary to wear a mask. And even more serious, that open spaces or empty environments do not present risk of contagion. We watch daily the highest health authorities conducting interviews in public without a mask, just keeping a distance of two meters between themselves. As we have seen, potentially contaminating droplets remain suspended in the air for several hours, and even days after the environment has been visited by an infected person.

Of course, just the existence of droplets suspended in the air is not proof that this is a means of effective contamination, as they may not contain active viruses. However, as if the dramatic empirical evidence from the Amoy Gardens and South Korea (2015) cases were not enough, in addition to the current mass contamination in choral and religious cults, unequivocal experimental demonstrations of the effective transport of SARS-COV-2 by air have been published recently. A relevant study issued in the journal Nature showed the results of the air investigation in several environments of two hospitals in Wuhan dedicated only to patients infected with Covid-19. The analysis of aerosols collected from the air showed unequivocally the existence of the RNA of the SARS-COV-2 in several closed environments and, also, in two open places where there was a great movement of contaminated people [18]. Another paper published this year analyzed the air at Nebraska Hospital Center and also found the SARS-COV-2 in most environments occupied by patients with mild and moderate infections [19]. In these two studies it was not possible to confirm the existence of an active viruses. However, this doubt was finally resolved by a study published last March in the New England Journal of Medicine, where the presence of active SARS-COV-2 in droplets was observed more than three hours after they were artificially produced in the laboratory [20]. Actually, they can survive longer, because 3 hours was the maximum measurement time of that experiment. Viral stability was studied in aerosols with a diameter of less than 5 $\mu$m, kept in an environment of 65% relative humidity and temperatures between 21-23ºC, containing SARS-COV-2 in the initial concentration of $10^{5.25}$ TCID50/mL (Median Tissue Culture Infectious Dose/mL). During the 3 hours studied, the experiment found an exponential reduction in infectious titer from $10^{3.5}$ TCID50/mL to $10^{2.7}$ TCID50/mL per liter of air, resulting in an average life of 1.1 hours. Since the chemical composition of the studied droplets was not exactly the composition of the droplets expired by an infected person, these times may be slightly longer or shorter in a real environment. In addition, atmospheric conditions can also change these values. But, in any case, it is certain that under normal day-to-day conditions SARS-COV-2 remains active for several hours in the droplets suspended in the air.

Therefore, from an epidemiological point of view, there is also an essential physical question to be answered: how long can these infected drops remain suspended in the air? In principle, they have weight and should be deposited on the soil after some time. However, as demonstrated by the videos cited above, and differently from what important health institutions believe [21], small drops exhaled by people during breathing, speech, sneezing, coughing and sports activities can remain for several hours in the air and occupy large environments in minutes. Thus, understanding the physical processes involved in these phenomena is essential, both for making crucial decisions by those responsible for health policies, as well as for alerting people about the hidden risks in the environments in which they live. This work will try to provide safe elements, through theoretical calculations, about some physical processes responsible for the transmission of the virus.

**Production and dispersion of droplets in the air**

The dynamic process involved in coughing, sneezing, or speech that exhale and disperse drops from the mouth and respiratory system is very complex. An exact physical calculation of this process is practically impossible, since it involves atmospheric science, aerosol physics, mechanics, thermodynamics, fluid mechanics, biochemistry, etc. The process can be summarized as follows. Approximately two liters of air, abruptly expelled from the mouth with speed 42 km/h in cough (14 km/h in speech) [22], create a turbulent flow of great penetration into the ambient air. Experiments have shown that its flow lines can infiltrate more than 2 meters into ambient air before they enter dynamic equilibrium. The ejected air is supersaturated with water vapor, has the temperature of the human body, and carries thousands of drops of saliva and/or secretions from the pulmonary tract. According to the recent measures mentioned above, these drops have diameters between fractions of micrometers up to fractions of centimeters and are composed of water, glycoproteins, salts, several other organic compounds, and, eventually, tens or thousands of viruses [23-25]. Once expelled, numerous and complex interactions between the hot air jet and ambient air begin, seeking thermodynamic equilibrium [26].

The droplets, depending on their compositions and air humidity, can evaporate or absorb water from the environment, and this process is fundamental to the survival of the virus in the environment. Unfortunately, there are no studies on the evaporation process of these drops of secretions taking into account their complete composition, and, in particular, how a colony of viruses carried by them affects their evaporation process. The few studies in this regard generally consider only the process of evaporation of drops of pure water mixed with salts, which is very far from the drops and actual mucous secretions that are expelled by the respiratory system. These studies showed that, in general, in environments with relative humidity above 60%, approximately, the drops absorb more than evaporate water into the air [27,28]. Therefore, in regions with high air humidity, it is expected that the survival of the virus in the drops is greater. In fact, in the current COVID-19 pandemic in Brazil, the Amazonian states that house the forest have presented transmission rates higher than 20%, while in southern states this rate has been less than 1%. During the first two months of the outbreak, in the regions surrounding the forest, the average temperatures were always above 30ºC and the average relative humidity, well above 70%. In southern states, average temperatures and average relative humidity were not higher than 22ºC and 40%, respectively. Therefore, apparently, there is no direct relationship between high ambient temperature and decreased transmissibility. On the other hand, air humidity seems to be a primary factor in the development of the epidemic. To better understand this relationship, it is interesting to know the amount of water vapor that actually exists in the atmosphere, that is, its absolute humidity. One kilo of air with relative humidity 70%, at 30ºC, contains approximately 19g of water in the form of vapor, while at relative humidity of 40%, at 22ºC, the amount of water is only 6g. That is, the process of water evaporation is very difficult in the Amazonian weather conditions, and, in fact, the drops expelled by an infected person will absorb water from the atmosphere, allowing the viruses to survive much longer in suspension or deposited on surfaces. This physical process may be important in explaining the difference observed in the transmissibility rates observed in Brazil. However, a detailed epidemiologic study, which takes into account the social and economic differences between these states, needs to be carried out to confirm this indication. In any case, it is unlikely that only the socioeconomic factors explain such a difference in transmissibility between these regions.

Finally, after these processes of production and evaporation of the drop, its equilibrium size will define its dynamics and, of course, the progression of the epidemic according to three scenarios of contagion: 1) the drop can be transmitted directly to another person and infect their mucous membranes and/or conjunctivae, in the so-called drop transmission mode; 2) it can succumb to the force of gravity and deposit itself dangerously on the ground, on furniture and objects, in the transmission mode by hand contact; and/or 3) in the case of smaller droplets that evaporate, reduce in size and form the so-called droplet nuclei, which are extremely small and can remain suspended for hours or days in the air, still containing water, salts, organic compounds and the colony of viruses. These smaller droplets and/or the droplet cores are responsible for the so-called airborne transmission. Although the relative importance of these three forms of contagion has been debated for years during the epidemics of the Influenza and corona viruses, definitive conclusions have not yet been reached due to the lack of detailed studies on airborne transmission. In an attempt to help to understand this process, we will calculate the dynamics of these drops in two situations: first, supposing that they fall into the air at rest; and then the droplets moving in an environment with a constant and vertical airflow upwards which can be produced by an air conditioning system and/or air renewal. From an epidemiological point of view, the most relevant physical variable is the time of permanence of the droplets in the air.

**Theoretical calculations**

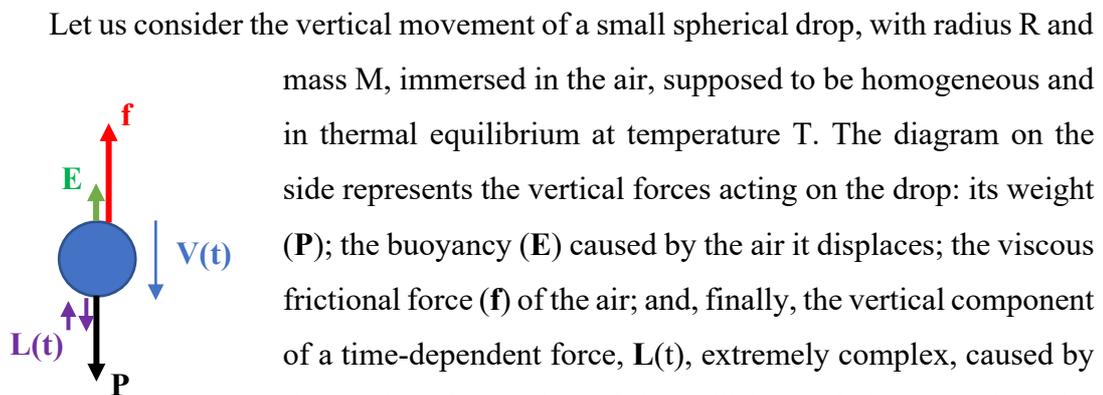

Let us consider the vertical movement of a small spherical drop, with radius R and mass M, immersed in the air, supposed to be homogeneous and in thermal equilibrium at temperature T. The diagram on the side represents the vertical forces acting on the drop: its weight (**P**); the buoyancy (**E**) caused by the air it displaces; the viscous frictional force (**f**) of the air; and, finally, the vertical component of a time-dependent force, **L**(t), extremely complex, caused by the random fluctuation of the collisions of air molecules with the drop, called Langevin force. This last force is significant only for very small particles when statistical variations in air density can cause macroscopic displacements of the drop, the so-called Brownian motion. This stochastic force varies very quickly over time and has a mean value of zero if it is calculated over a large time interval. The differential equation that governs the velocity of the drop's fall will be:

$$M\frac{dV}{dt} = L(t) + f - (P - E) \qquad (1)$$

As we are interested in low speeds, the viscous frictional force can be approximated by $f = -\mu V$, where μ is the viscous friction coefficient of air, given by Stokes' law, which, in the case of a sphere of radius R, will be μ = 6πηR. The air viscosity coefficient, η, depends on the ambient temperature, atmospheric pressure and drop radius. The exact solution of Eq. 1 is very complex due to the random characteristics of the Langevin force. However, for our evaluation, at first approximation we can disregard its effect and discuss it later. In this case, the solution is very simple, and the drop that starts from rest will fall with the velocity:

$$V(t) = \frac{(P-E)}{\mu}\left[1 - \exp\left(-\frac{t}{\tau}\right)\right] \qquad (2)$$

where $\tau = \frac{M}{\mu} = \frac{2\rho R^2}{9\eta}$ is the constant that governs the temporal behavior of the velocity of the drops, and ρ is the density of the substance that constitutes them. Equation 2 shows that their movement is extremely dependent on their sizes. To analyze the epidemiological implications of this expression, we will calculate the fall of drops of human saliva in the air at a pressure of 700 mmHg and a temperature of 25º C. Table 1 shows the values of the time constant τ calculated for drops of radius ranging from 0.1μm to 50 μm.

Table 1. Fall of saliva droplets with radius R, where τ are their time constants, $V_{lim}$ are their final velocities and $T_{tot}$ are the total time they take to travel the height of 1.80 meters.

| R (μm) | τ (μseg) | $V_{lim}$ (mm/s) | $T_{tot}$ |
|---|---|---|---|
| 0.1 | 0.23 | 0.0022 | 227 hours |
| 0.5 | 3.55 | 0.035 | 14.3 hours |
| 1.0 | 13.1 | 0.13 | 3.8 hours |
| 10.0 | 1.3x10³ | 11.9 | 2.5 min |
| 50.0 | 3.0x10⁴ | 296.3 | 6.1 seg |

On the other hand, Figure 1 shows the dynamics of two drops, of radii 0.5 μm and 1.0 μm, where are shown the variations of their velocities with time. We observed that the drops start to fall from rest, increase their velocities, and very soon acquire constant velocities, $V_{lim}$. What is surprising about these results is the order of magnitude of the times and speeds: in a few microseconds, the smallest drops start to fall with constant and extremely small speeds. For example, in this ideal situation, with stationary air, a drop of radius 0.1 μm falls with speed 2.2 μm/s, and a drop of radius 0.5 μm falls with speed 35 μm/s, approximately. Evidently, these values are negligible when compared to the speeds of the random air currents that exist in real environments, which are in the order of cm/s.

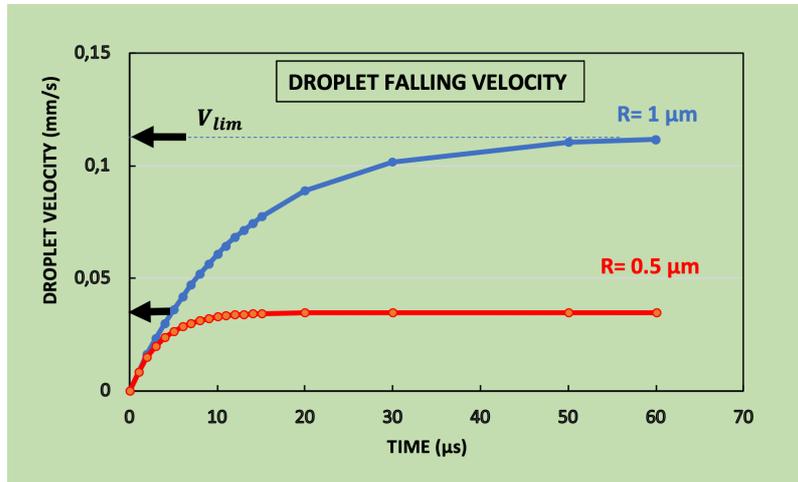

Figure 1. Velocities of two droplets of saliva, radius 0.5 μm and 1.0 μm, which fall into the air from rest.

The last column of Table 1 shows the time, $T_{tot}$, that the drops take to reach the ground starting from a height of 1.80 meters, remembering that these values were evaluated disregarding the Langevin force. Let us now evaluate, separately, the effect of Brownian motion in one direction. Since it is produced by a random force, which has the same probability of acting in any direction, the particle has the same probability of moving up and down. Therefore, the average displacement, calculated over a large time, will always be zero. However, Einstein showed that this is not the case for the mean square displacement, whose value can be calculated by $y_{rms} = \sqrt{2\,Dt}$, where D is the diffusion coefficient given by the Einstein relation, D= kT/μ, where k is the Boltzmann constant. When applied to our 0.1 μm radius drops, these equations provide dispersions in the position of only 0.1 mm per hour. That is, we can disregard the Brownian effect in our evaluation. In fact, the complete solution of the Eq. 1 was obtained by Saka et al. [29] using the Fokker-Planck equation for the probability density of the vertical position of small spheres. The theoretical solution found was compared with experimental measurements of the temporal evolution of the spatial distribution of small latex spheres of a diameter of 1 μm suspended in water. The agreement was excellent. However, when we perform our calculations for this same system disregarding the Langevin force, we find a good approximation for the temporal evolution of the mean value of the positions of the latex spheres. This is an experimental confirmation that we can ignore the stochastic force in our evaluations.

**Effect of air conditioners and heating systems**

Even considering the air at rest, the last column of Table 1 shows that drops of diameters 0.2 μm, 1.0 μm, and 2.0 μm would remain suspended for several hours in the air. However, in real situations, the air is never at rest, there are internal currents produced by differences in temperature and pressure [19], by the movement of objects or people around the environment and, mainly, by air conditioning and heating systems. These currents are much larger than those calculated above and drag small suspended particles, being the main responsible for the movement of dust grains observed by Lucretius, and also for the fluctuation and movement of microdrops filmed in the experiments described above. These currents are even more important when the environment has an air renewal and/or conditioning system that creates a continuous flow of air in a more or less fixed direction. We will evaluate this phenomenon in the case of an aspiration system placed on the ceiling of the environment, producing an airflow in the vertically upward direction, with speed **V**. In this case, the viscous friction force of the air passing through the droplet can compensate for and/or overcome the weight force, causing it to remain stopped and/or be aspirated towards the ceiling. Let's consider the equilibrium limit case, when friction exactly compensates for the weight of the droplet and it stays at rest at a certain height of the ground. The diagram on the side shows the forces acting on it. Disregarding Langevin's force, the speed of air that holds a drop at rest can be easily calculated for its different sizes. Figure 2 shows the air velocity necessary to balance drops of different radii, where we observe that they are very small velocities, of a few mm/s, even for the largest drops considered. That is, if the air conditioning system is not well dimensioned, it eliminates the smallest drops, but it can keep larger drops in suspension and/or drastically decrease its fall times, precisely the drops that have a greater potential for infection because they can carry a greater amount of viruses.

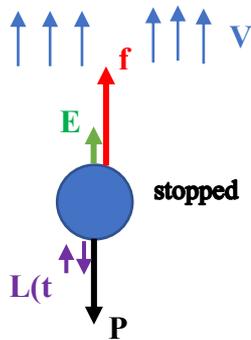

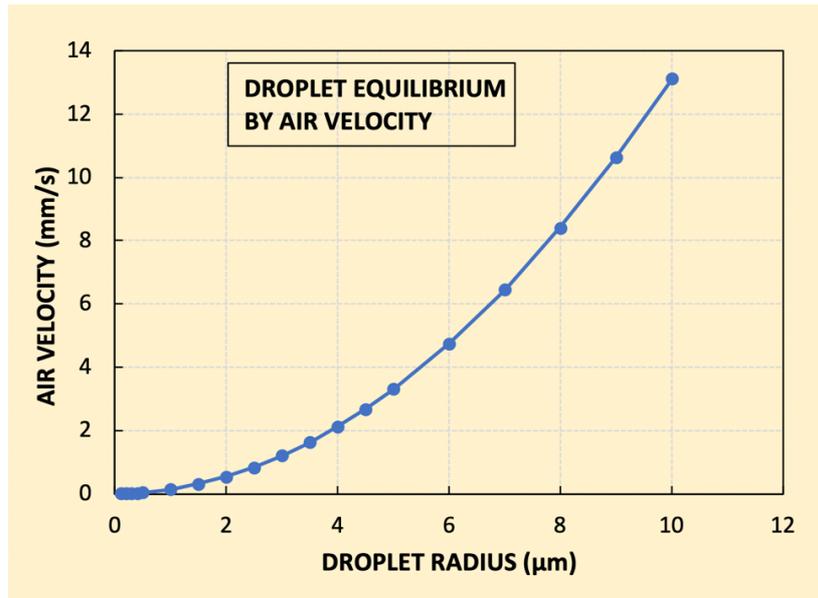

Figure 2. Vertical airflow velocity required to keep drops of saliva suspended in the air at rest.

For example, a continuous vertical flow of air with a speed of only 3 mm/s, approximately, keeps drops of diameter 10 μm suspended at rest for an unlimited time. Besides, larger droplets would have very slow fall speeds. It is important to remember that the coronavirus has a more or less spherical shape with a diameter of the order of nanometers, that is, the drops we dealt with in this work can carry from hundreds to millions of viruses. The crucial question from an epidemiological point of view is whether these small drops carry active viruses after the evaporation processes and how long this virus survives in these droplets. As mentioned earlier, recent experience has found the significant presence of SARS-COV-2 active in drops that have been suspended in the air for more than three hours. This same study measured the survival time of the coronavirus on various surfaces, finding that the they can remain active for many hours and even days outside the human body.

Finally, we must discuss the effect of heating systems placed on the ground, as they also create an upward airflow since this is the natural direction of the heated air. Our calculations have demonstrated that these airflows can avoid contaminated droplets from settling on the floor, which can stay floating in the air for a long time, increasing the risk of contamination in the environment. Perhaps, this may be one of the factors that increase the spread of viral epidemics during the winter.

**Summary, conclusions and warnings**

There is still much debate about the main forms of contamination by the coronavirus. The speed and extent of the current outbreak called into question the dominant conception in official health entities that believed in the preponderance of direct transmission. As we have seen, recent experiences have filmed the movement of drops produced by speech, coughing and sneezing, revealing the potential for viral contamination by air. In addition, studies that analyzed the air of Nebraska and Wuhan hospitals confirmed the presence of SARS-COV-2 in air-suspended droplets and micro droplets in various environments, without however ensuring that such viruses were active. But another recent experiment completed that study by confirming the presence of SARS-COV-2 active in droplets in the air, even several hours after they were released into the atmosphere. Therefore, this set of works has shown that airborne contagion by SARS-COV-2 can occur long after an infected person has spoken, coughed or sneezed in an environment. These experiments also showed that this contamination can occur several meters away from the original virus source. All this knowledge now accumulated could explain the rapid and widespread of the current outbreak of coronavirus, since contagion by air was ignored by health authorities in the first months of the pandemic. The general use of a mask by the population was only recommended late. This experimental data also strengthens the hypotheses about the importance of airborne infection in the coronavirus epidemics of 2002 in Hong Kong and 2015 in South Korea.

From a theoretical point of view, the results of the calculations that we present in this work explained the dynamic behavior of the drops observed in the videos mentioned before and are extremely important from an epidemiological point of view, as they proved that the smallest drops can remain in the air for several hours after they have been produced by some contaminated person. That is, long after that person has left the environment, or passed through a public place, the air remains dangerously contaminating. Of course, the probability of contamination depends on both the concentration of droplets in the air and the concentration of people in the place. The argument that the wind disperses the drops has made people feel more protected in open places. However, the same wind that can disperse the drops can also carry them and project them on passersby. It should be remembered that a single of these drops inhaled or deposited in the mouth or eyes can contaminate another person, whether on the beach, on the street, in the elevator, at home or on public transport. Today, to avoid such contagions, continuous use of the mask by the population is recommended. However, it is surprising that the use of eye protection remains mandatory only for health professionals. Extensive study published on June 1, 2020 in The Lancet journal, analyzing empirical data from 16 countries on 6 continents, concluded that the probability of infection by SARS-CoV-2 decreases by 10.6% when using a protection for the eyes [20]. That is, the risk of contagion through the eyes is very high and continues to be minimized by health authorities, including the World Health Organization, as had happened in the case of masks. This may be a new mistake in combating the pandemic.

Another important result of our calculations and the videos mentioned above refers to the safety distance recommended by all health authorities. Once again, the recommendation of two meters of distance is based only on direct virus transmission, minimizing the airway mode, although there are no studies to prove this hypothesis. As we have seen, a few minutes after being expelled, the contaminated droplets spread quickly and occupy environments over 8 meters long, remaining for several hours floating in the air. Therefore, important warnings must be made: the distance of 2 meters is not safe for those who do not wear a mask and, if we consider the infection by the eyes, the distance of 2 meters is not safe even for those wearing a mask. This may explain the large number of infected people who wore masks frequently. An interesting study that could be done concerns the frequency of contamination in patients who wear corrective glasses compared to those who do not, because, although these glasses are not ideal to protect against the virus, they do offer a little protection to the eyes. In addition, it should not be forgotten that drops in suspension can also be deposited on our face, hair and clothes. This explains that one of the highest concentrations of infected droplets in the air of Wuhan hospitals was measured precisely in the area where health workers removed their protective clothing. Where, moreover, there may have been a great contamination of these professionals.

The SARS-CoV-2 pandemic in Brazil will provide much information for epidemiological studies. For example, apparently, there is no direct relationship between high ambient temperature and decreased transmissibility, which refutes the hope that the outbreak could be reduced during the summer in the northern hemisphere. In the first three months of the current COVID-19 pandemic in Brazil, Amazonian states, with average temperatures above 30ºC, had virus transmission rates higher than 20%, while in southern states this rate was lower than 1%, although their average temperatures did not exceed 22ºC. Another relevant conclusion that can be obtained from the SARS-CoV-2 outbreak in Brazil concerns the apparently relevant role of air humidity. In the regions surrounding the Amazon rainforest, the average relative humidity was well above 70%, while in southern states the average relative humidity was not higher than 40%. Therefore, air humidity seems to be a major climatic factor in the development of the epidemic. A possible explanation for this phenomenon would be the fact that the drops expelled by an infected patient do not evaporate in relative humidity greater than 60%, but instead absorb water from the air, which allows viruses to survive much longer in suspension or deposited on surfaces. However, only a detailed epidemiologic study, also considering the socioeconomic differences between the North and South regions of Brazil, can confirm these hypotheses.

Finally, another important alert that our calculations recommend concerns about air conditioning and heating systems. As we have demonstrated, if poorly positioned and/or sized, they can keep larger droplets in suspension for much longer than they would naturally be, as they create air flows that support the weight of these drops and keep them in equilibrium. Ideally, from this point of view, the air would be aspirated near the ground or at least laterally, with enough flow to clean the environment. In addition, it is essential that the air is renewed with fresh outside air. Otherwise, if the device only circulates ambient air, and if its filter is not able to block microparticles, it will work as a dangerous virus spreader, as was the case with the restaurant that started the new coronavirus epidemic in Guangzhou, in 2020. On the other hand, in the absence of air renewal systems, two windows must be opened at the same time to create a flow out of the residence. Otherwise, droplets are only transported from one environment to another. In all countries, health authorities recommend that slightly contaminated patients stay at home. In such cases, special care should be taken with the ventilation of the environments to prevent contaminated air from the patient's room from being transported to the others. In addition, it is recommended to wear a mask and goggles for the other inhabitants of these residences.